\documentstyle[12pt]{article}
\advance \textheight by \topskip
\begin{document}

\begin{titlepage}
\hbox to \hsize{\hfil hep-th/9403118}
\hbox to \hsize{\hfil IHEP 94--21}
\hbox to \hsize{\hfil March, 1994}
\vfill
\large \bf
\begin{center}
LOCAL SOLUTIONS OF HARMONICAL AND BI-HARMONICAL EQUATIONS, UNIVERSAL FIELD
EQUATION AND SELF--DUAL CONFIGURATIONS OF YANG--MILLS FIELDS IN FOUR
DIMENSIONS
\end{center}
\vskip 1cm
\normalsize
\begin{center}
{\bf A. N. Leznov\footnote{E--mail: leznov@mx.ihep.su}}\\
{\small Institute for High Energy Physics, 142284 Protvino, Moscow Region,
Russia}
\end{center}
\vskip 2.cm
\begin{abstract}
\noindent
A general method for the construction of solutions of the d'Alambertian
and double d'Alambertian (harmonic and bi--harmonic)  equations with local
dependence of arbitrary functions upon two independent arguments is
proposed.  The connection between solutions of this kind and self--dual
configurations of gauge fields having no singularities is established.
\end{abstract}
\vfill
\end{titlepage}

\section{Introduction}

The general solution of the d'Alambertian equation in $D$-dimensional space
depends upon two arbitrary functions of $(D-1)$ independent arguments. In
the space of two dimensions the general solution of d'Alambert's equation
$\phi_{\xi,\bar \xi}=0$ has the form $\phi=f(\xi)+\bar f(\bar \xi)$
and locally depends upon two arbitrary functions $f$ and $\bar f$.
In spaces of higher dimensions the general solution may be expressed
either as an integral transform, or else by using what is much the same thing,
a construction using Greens functions where the dependence on arbitrary
functions becomes nonlocal. Nevertheless in some problems as for example in
the theory of radiation  some partial solutions arise with a local dependence
of the solution on arbitrary functions (the number of independent
arguments being less  than  is necessary to solve the general Cauchy
problem). We demonstrate this situation with  the well--known example
of  in and out--going waves  in the case($D=4$)
\begin{equation}
\Phi={f(r+it,\exp(-i\phi) \cot {\theta \over 2})\over r}+{\bar f(r-it,
\exp(-i\phi) \tan {\theta \over 2})\over r}\label{1}
\end{equation}
where $ t=x_4,$ $r=\sqrt{x_1^2+x_2^2+x_3^2}$, $x_1=r\sin \theta\sin \phi,
x_2=r\sin \theta\cos \phi, x_3=r\cos \theta$; $f,\bar f$ are
arbitrary functions of their two respective arguments . We have exhibited
(\ref{1}) only for illustrative purposes.

The goal of this paper is to explain a method of construction of  solutions
of this kind for the d'Alambert and double d'Alambert equations in three and
four-dimensional spaces.

\section{The general construction}

We shall employ the following notation; $y=x_1+ix_2, z=x_3+ix_4$
($\bar y=y*,\bar z=z*$); $\framebox(10,10){};\ \Psi=\Psi_{y,\bar y}+\Psi{z,
\bar z}$.

Let us introduce three variables $ u_1=y+\lambda \bar z,u_2=z-\lambda \bar y,
u_3=\lambda$ and consider the equation $ F( u_1,u_2,u_3)={\rm Constant} $
as an implicit definition of the function $\lambda$ as a function
of independent arguments $ y,\bar y,z, \bar z$; $ F $ is an
arbitrary function of its three arguments. From these
definitions it follows immediately that:
\begin{equation}
\lambda_y=-{F_1\over F_{\lambda}}\quad\lambda_{\bar z}=-\lambda{F_1
\over F_{\lambda}},\quad\lambda_z=-{F_2\over F_{\lambda}},\quad
\lambda_{\bar y}=\lambda{F_2\over F_{\lambda}}\label{2}
\end{equation}
where $F_i=F_{u_i},F_{\lambda}\equiv \bar z F_1-\bar y F_2+F_3$. Or
\begin{equation}
\lambda_{\bar z}=\lambda \lambda_y,\quad\lambda_{\bar y}=-\lambda \lambda_z
\label{3}
\end{equation}
In the case of one pair of variables (say $y,\bar z$) (\ref{3})
is known as equation of Monge \cite{1} who first  found its exact
solution in implicit form. In recent years the generalisation of
this equation on the space of higher dimensions has been intensively
investigated in a series of papers by D.Fairlie and his collaborators\cite{2}
who call this generalised equation the Universal Field Equation.

It is easy to check by direct calculations that each function $\alpha\equiv
\alpha(u_1,u_2,u_3)$ is annihilated by the pair of operators $D_1\equiv
\partial_{\bar z}-\lambda\partial_y,\quad D_2\equiv\partial_{\bar y}+
\lambda\partial_z$. From this fact and definitions above the following
proposition results.

Let
\begin{equation}
\phi^s=\alpha (u_1,u_2,u_3) F^s_{\lambda} \label{4}
\end{equation}
then
\begin{equation}
(\partial_{\bar z}-\lambda\partial_y) \phi^s=-s\lambda_y\phi^s,\quad
(\partial_{\bar y}+\lambda \partial_z) \phi^s=s\lambda_z\phi^s\label{5}
\end{equation}
and
\begin{equation}
\framebox(10,10){}\;\phi^s=(s+1)({\lambda_z\phi^s_y-\lambda_y\phi^s_z})
\label{6}
\end{equation}
As a corollary in the case $s=-1$ we obtain
\begin{equation}
\phi^{-1}_{\bar z}=(\lambda \phi^{-1})_y,\quad\phi^{-1}_{\bar y}=
-(\lambda \phi^{-1})_z,\quad \framebox(10,10){}\;\phi^{-1}=0\label{7}
\end{equation}.

So with the help of an arbitrary root of the equation $F(u_1,u_2,u_3)={\rm
constant}$ it is possible to construct the function $\phi^{-1}$ which is an
exact the local solution of d'Alambert equation.

\section{Simple examples}

In the case when $F$ is a polynomial of second order with
respect to $\lambda$ we have $F-{\rm constant}=A\lambda^2+B\lambda +C=0$.
$$
\lambda_{1,2}={-B\pm \sqrt{B^2-4AC}\over 2A},\quad F_{\lambda}=2A\lambda
+B,
$$
$$
\phi^{-1}_{1,2}={1\over 2A\lambda_{1,2}+B}=\pm{1\over A(\lambda_1-
\lambda_2)}=\pm{1\over \sqrt{B^2-4AC}}
$$

Let $F=u_1u_2+a\lambda^2+a_0\lambda+\bar a=(yz+\bar a)+\lambda
(z\bar z-y\bar y)+\lambda^2(\bar z\bar y+a)=0$ and for the corresponding
solution
of the d'Alambert equation we obtain
$$
\phi={1\over \sqrt {{(z\bar z-y\bar y+a_0)\over 4}+(yz+\bar a)
(\bar z\bar y+a)}}
$$
If $a=\bar a=a_0=0$ then we have the usual $O(4)$ invariant solution
$\phi={1\over z\bar z+y\bar y}$ with singularity in the origin
of coordinates. In the general case the singularities of this
solution are lying an the curves $ y=-{a\over z},\bar y=-{a\over
\bar z}, (z\bar z)^2+a_0(z\bar z))-a\bar a=0$.

The choice $F$ in the form $F=u_1+u_3u_2\equiv y+\lambda(z+\bar z)-\lambda^2
\bar y$ exactly corresponds to the solution of (\ref{1}). In this
section we have assumed a reality condition  which will be given in explicit
form in section 6.

\section{The three-dimensional case}

The solutions of three-dimensional Laplace equation may be
obtained from construction  described above by  special choice
of the initial functions $F$ and $\alpha$
\begin{equation}
F=F(u_1+u_3u_2,u_3),\quad \alpha(u_1+u_3u_2,\lambda)\label{8}
\end{equation}
( compare with the last example of the previous section).
 Indeed in this case $u_1+u_3u_2\equiv y+\lambda(z+\bar z)-\lambda^2
\bar y$ and $\lambda$ become independent of the argument $z-\bar z$.
For all functions of the above construction $\partial_z=\partial_{\bar z}$
and the d'Alambertian operator goes into three-dimensional Laplace one.

\section{Bi d'Alambertian equation}

By techniques of section 2 it is not difficult to prove the
following relation
$$
\framebox(10,10){}\; \framebox(10,10){}\; \mbox{ln} \phi^s=-s
\framebox(10,10){}\; \framebox(10,10){}\; \mbox{ln} F_{\lambda}=
$$
\begin{equation}
{}\label{9}
\end{equation}
$$
s(\frac{\partial^2 \lambda}{\partial y^2}\frac{\partial^2 \lambda}
{\partial z^2}-(\frac{\partial^2 \lambda}{\partial y \partial z})^2)=
s{D_4(F)\over F^4_{\lambda}}
$$
where
$$
D_4(F)=Det\left(
\begin{array}{cccc}
 0  &    F_1 &    F_2 &    F_3 \\
F_1 & F_{11} & F_{12} & F_{13} \\
F_2 & F_{21} & F_{22} & F_{23} \\
F_3 & F_{31} & F_{32} & F_{33}
\end{array}
\right)
$$

(It follows from the linear equation for $\phi^0$
(\ref{6}) that each function of three  (exactly two) arguments $u_1,u_2,u_3$
satisfy the bi d'Alambertian equation).

The condition $\framebox(10,10){}\; \framebox(10,10){}\; \mbox{ln} F_{\lambda}=
0$ means that either $D_4(F)=0$ or it must be divided by $ F-{\rm constant}$
because $F={\rm constant}$ is the equation which determines $\lambda$.

The equation $D_4(F)=0$ is exactly the Universal equation of D.B.Fairlie et al
in space of three dimensions. The general solution of it is constructed in
implicit form  and some explicit solutions are noted in \cite{3}.

So if we take $F$ as the solution of Universal equation in the
space of three dimensions then function
\begin{equation}
\phi=\alpha (u_1,u_2,u_3)+ \ln F_{\lambda} \label{10}
\end{equation}
will be a local solution of the Bi d'Alambertian, or bi-harmonic  equation.

\section{The condition of reality}

The reality conditions on the solutions ${1\over F_{\lambda}}, \ln
F_{\lambda}$ constructed here give further restrictions on the choice of
initial function $F$. Namely
$$
F^*(u_1,u_2,u_3)=F(-{u_2\over
u_3},{u_1\over u_3},-{1\over u_3})
$$
In examples of section 3 we have used this restriction without any special
mention of it.

By direct calculations it is easy to convince that the Universal
equation in three dimensions is invariant with respect to
discrete substitution
$$
u_1\to -{u_2\over u_3},\quad u_2\to {u_1\over u_3},\quad
u_3\to -{1\over u_3}
$$
(this is the partial case of $SL(4,R)$ transformation with respect to
which is invariant Universal equation in three dimensions \cite{4}). From
the last invariance it follows that among the solutions of Universal
equation there are those for which condition of reality is satisfied.

\section{Connection with the self-dual configurations of
Yang-Mills field theory}

We shall explain this connection on the well-known example of t'Hoof's
solution. All dynamical variables in this case are expressed in terms of
single function $\phi$ -the solution of d'Alamber equation in four
dimensions. The topological charge density has the form
$$
q=\framebox(10,10){}\; \framebox(10,10){}\; \phi
$$
Let us choose $\phi$ in the form
$$
\phi=1+\sum \phi_{\nu}^{-1}
$$
where each $\phi_{\nu}^{-1}$ is the local solution of d'Alamber
equation of section 2 (\ref{7}). If we want that singularities of
$\phi_{\nu}^{-1}$ ( the zeroes of $F_{\lambda}$) will not give
a contribution into the topological charge density it is necessary to demand
$$
 \framebox(10,10){}\; \framebox(10,10){}\; \mbox{ln} F_{\lambda}=0.
$$
So $F$ can not be taken arbitrary but only as solution of
Universal equation (see section  5).

The well-known 5n-th parametrical instanton solution arise
under the choice $\phi_{\nu}^{-1}$ in the pole form $
\phi_{\nu}^{-1} ={1\over (x-a)^2}$. In this
case $ \framebox(10,10){}\; \framebox(10,10){}\; \mbox{ln}{}\; (x-a)^2=0$.
This means that singularities in each pole may be canceled by appropriate
gauge transformation.

The analogue consideration may be applied to self-dual solution
which arise by the formalism of discrete transformation \cite{5}.

\section{Conclusion remarks}

The main results of the present note are containing in formulae (\ref{7}),
(\ref{9}) which give the local solutions of harmonical and bi-harmonical
equations in the terms of arbitrary function of third variables
and its derivatives in the first case and in terms of solution of Universal
equation in three dimensions in the second. The most unexpected
is the fact of correlation between the solutions of Universal and
bi-harmonical equations in the problems of self-dual
configuration of Yang-Mills gauge theory.

The author is indebted to D.B.Fairlie for discussion of
results of this paper.


\begin{thebibliography}{**}

\small

\bibitem{1}
D. Zwilinger {\it Handbook of differential equations} (New York,
Acad. Press., 1989).

\bibitem{2}
D.B. Fairlie, J.Govaerts and A. Morzov {\it Nuclear Phys.} B 373
(1992), 214-232.

\bibitem{3}
D.B. Fairlie and J. Govaerts {\it Linearisation of Universal Field equations}
{DTP-92/47, NI-92/011, December 1992}.

\bibitem{4}
V.B.Deryagin and A.N. Leznov, {Geometrical symmetries of Universal
equation},(in preparation)

\bibitem{5}
A.N. Leznov, {Nonlinear symmetries of integrable systems}
{\it J.of Sov.Lazer. Research} {\bf 3-4}, 278-288, (1992)

Ch. Devchand and A.N.Leznov, Backlund transformation for supper-symmetric
self-dual theories for semisimple gauge groups and hierarchy of $A_1$
solutions. Preprint IHEP DTP 92-170, (1992) ( to be published in
Commun. Math. Phys).
\end{thebibliography}
\end{document}